\documentclass[12pt]{article}

\catcode`\@=11

\global\arraycolsep=2pt
\usepackage{amsbsy,amssymb,latexsym,amsfonts,amsmath}
\usepackage{graphicx,color}

\begin{document}
\begin{flushright}
\parbox{4.2cm}
{RUP-19-13}
\end{flushright}

\vspace*{0.7cm}

\begin{center}
{ \Large Holographic Dual of Conformal Field Theories 
with Very Special $T\bar{J}$ Deformations}
\vspace*{1.5cm}\\
{Yu Nakayama}
\end{center}
\vspace*{1.0cm}
\begin{center}

Department of Physics, Rikkyo University, Toshima, Tokyo 171-8501, Japan

\vspace{3.8cm}
\end{center}

\begin{abstract}
Very special $T\bar{J}$ deformations of a conformal field theory are irrelevant  deformations that break the Lorentz symmetry but preserve the twisted Lorentz symmetry. We construct a holographic description of very special $T\bar{J}$ deformations. We give a holographic recipe to study the double trace as well as single trace deformations. The former is obtained from the change of the boundary condition while the latter is obtained from the change of the supergravity background. 

\end{abstract}

\thispagestyle{empty} 

\setcounter{page}{0}

\newpage




\section{Introduction}

Inspired by exactly marginal and integrable $J\bar{J}$ deformations (e.g. see \cite{Chaudhuri:1988qb,Hassan:1992gi,Forste:1994wp,Forste:2003km,Georgiou:2019jws}) and irrelevant but integrable $T\bar{T}$ deformations of two-dimensional conformal field theories \cite{Zamolodchikov:2004ce,Smirnov:2016lqw,Cavaglia:2016oda,Dubovsky:2017cnj}, the Lorentz breaking, irrelevant but integrable $T\bar{J}$ deformations of conformal field theories  have been vigorously studied in recent works \cite{Guica:2017lia,Bzowski:2018pcy,Chakraborty:2018vja,Apolo:2018qpq,Aharony:2018ics,Nakayama:2018ujt,Araujo:2018rho,Sun:2019ijq,Guica:2019vnb,Giveon:2019fgr,LeFloch:2019rut,Conti:2019dxg,Chakraborty:2019mdf}. Better understanding of ultraviolet completions of power-counting non-renormalizable theories is theoretically important, in particular with Lorentz violations, because they might give a hint toward (Lorentz violating) ultraviolet completions of quantum gravity in higher dimensions (such as Horava-Lifshitz gravity \cite{Horava:2009uw}), and we expect that $T\bar{J}$ deformations play a role of such toy models.

The generic $T\bar{J}$ deformations break the global conformal symmetries of $SL(2,\mathbb{R})\times SL(2,\mathbb{R})$ down to $SL(2,\mathbb{R}) \times \mathbb{R}$ (i.e. left-moving translation, left-moving dilatation, left-moving special conformal transformation, and right-moving translation). In the subsequent work \cite{Nakayama:2018ujt}, we have proposed the ``very special" types of $T\bar{J}$ deformations which preserve the additional (right-moving) twisted Lorentz transformation while still breaking the right-moving special conformal transformation. The four-dimensional analogue of the symmetry structure here is the so-called very special conformal symmetry (based on the earlier idea of very special relativity proposed by Cohen and Glashow \cite{Cohen:2006ky,Cohen:2006ir,Nakayama:2017eof,Nakayama:2018qcs,Nakayama:2018fib}) and it is considered very special (in addition to the original meaning of the very special relativity) and peculiar because  unitarity or locality is typically sacrificed for their realizations.

The success of field theoretic ultraviolet completions of $T\bar{J}$ deforamtions has naturally led to holographic studies of the $T\bar{J}$ deformations.\footnote{The analogous (holographic) studies of $T\bar{T}$ deformations can be found in \cite{Giveon:2017nie,Giveon:2017myj,Asrat:2017tzd,Giribet:2017imm,Kraus:2018xrn,Cottrell:2018skz,Baggio:2018gct,Babaro:2018cmq,Gorbenko:2018oov,Cardy:2018sdv,Dubovsky:2018bmo,Datta:2018thy,Aharony:2018bad,Bonelli:2018kik,Conti:2018tca,Baggio:2018rpv,SUSY,Song:2019txa,Jiang:2019tcq}.} By construction, the asymptotic behavior of the gravity should be different from that of the AdS space-time that is dual to the undeformed theory, and this may lead to novel types of holographic setup. Since we do not know the very definition of quantum gravity beyond the asymptotically AdS space-time, the holographic construction of the $T\bar{J}$ deformations  may give a new direction to understand the nature of quantum gravity in more general settings. 

There are two different types of holographic $T\bar{J}$ deformations. The one is given by the double trace $T\bar{J}$ deformations \cite{Bzowski:2018pcy}. This is a natural realization of the original field theoretic idea of $T\bar{J}$ deformations in the sense that we study the composite operators of $T$ and $\bar{J}$ as a double trace operator in the large $N$ setup. The construction is ubiquitous in every (holographic) conformal field theories with a conserved (chiral) current, and holographically realized by changing the boundary conditions of asymptotic infinity.
The other is the so-called single trace $T\bar{J}$ deformations \cite{Chakraborty:2018vja,Apolo:2018qpq}. This is given by the operator which has the same quantum number as the double trace $T\bar{J}$, but is a single trace operator. The existence of such operators is not necessarily guaranteed in generic holographic conformal field theories with a conserved current, but in many concrete examples of the AdS/CFT setups, the Kaluza-Klein towers of graviton play the role of such operators. The crucial difference of the single trace $T\bar{J}$ deformations is that the actual asymptotic behaviors of the metric (rather than the boundary conditions) are modified.

In this paper, we propose holographic descriptions of double trace as well as single trace very special $T\bar{J}$ deformations from effective field theory approaches in the bulk gravity. In dual conformal field theories, the very special $T\bar{J}$ deformations are accompanied by the twisted Lorentz symmetry \cite{Nakayama:2018ujt}, and we will implement the similar mechanism in holography. The holographic realizations are to some extent phenomenological and based on the effective bulk theories (rather than the full string embedding) because the realizations of non-compact symmetries in the bulk that we will need are not straightforward in the full string setup. It will be an interesting future direction to see if we may be able to embed our constructions in a full string theory background.

The organization of the paper is as follows. In section 2, we briefly review field theoretic aspects of very special $T\bar{J}$ deformations of two-dimensional conformal field theories. In section 3 we first construct holographic dual descriptions of double trace $T\bar{J}$ deformations and then we construct holographic dual descriptions of single trace $T\bar{J}$ deformations. In section 4, we conclude the paper with discussions.

\section{Very special $T\bar{J}$ deformation of conformal field theories}
Let us briefly review the very special $T\bar{J}$ deformations of two-dimensional conformal field theories proposed in \cite{Nakayama:2018ujt}. 
We use the light-cone notations and our convention is
\begin{align}
x^\pm = \frac{1}{\sqrt{2}}(t\pm x)
\end{align}
and we call the fields only dependent on $x^-$ as left-moving and the fields only dependent on $x^+$ as right-moving. For example, a component of the (traceless) energy-momentum tensor  $T^{+}_{\ -} = -T_{- -}$ satisfying $\partial_+ T^{+}_{\ -} = 0$ is left-moving. In the Euclidean setup $T_{--} = -T^{+}_{\ -}$ is identified with the holomorphic ($=$ left-moving in our convention) energy-momentum tensor $T(z)$ with $\bar{\partial} T(z) = 0$. Similarly a component of the chrial conserved current $J^{-}$ satisfying $\partial_- J^{-} = 0$ is identified with the right-moving or anti-holomorphic current $\bar{J}(\bar{z})$ with $\partial \bar{J}(\bar{z}) = 0$.

Suppose there is a conformal field theory with a (chiral) conserved current $\bar{J}(\bar{z})$. The $T\bar{J}$ deformation is formally defined as perturbing the conformal field theory by adding 
\begin{align}
\delta S =  \int d^2 z \mu  T \bar{J} \ 
\end{align}
to the action, where $T(z)$ is the holomorphic energy-momentum tensor of the undeformed theory.
This added term has the conformal dimensions of $(h,\bar{h}) = (2,1)$ and it is irrelevant with respect to the conventional dilatation $D= h + \bar{h}$, but we may study the ultraviolet completion. Note that the deformation breaks the Lorentz symmetry, but it preserves the right-moving conformal (or Virasoro) symmetry  together with the left-moving translation.

The idea of very special $T\bar{J}$ deformation is to further assume that the current $\bar{J}(z)$ is charged under a (non-compact) symmetry generator $K$:
\begin{align}
i[K, \bar{J}(z)] = \bar{J}(z)  \ .
\end{align}
Note that the charge is ``pure imaginary" (similarly to the ghost charge or dilatation charge because we assume $\bar{J}(z)$ is Hermitian) in contrast with the compact $U(1)$ symmetry like the electric charge. 
With a non-zero weight for $\bar{J}(\bar{z})$ under $K$, we may define the twisted Lorentz transformation $J_{\mathrm{twisted}} = L_0 - \bar{L}_0 + K$ such that the deformation is invariant under the twisted Lorentz symmetry. This is the idea of the very special $T\bar{J}$ deformation of conformal field theories.

Under the typical very special $T\bar{J}$ deformations, the left-moving special conformal symmetry is broken while the right-moving conformal symmetry is intact as can be seen from the study of the energy-momentum tensor (see \cite{Nakayama:2018ujt} for the detailed study). This is rather a generic feature of the twisted conformal field theories with deformations. The current $\bar{K}(\bar{z})$ associated with the symmetry generator $K$ plays the role of the Virial current.  

One may study correlation functions of very special $T\bar{J}$ deformed conformal field theories as studied in the $T\bar{J}$ deformed conformal field theories in \cite{Guica:2019vnb}. The additional feature is that we have the twisted Lorentz symmetry. We just emphasize here that the existence of the twisted Lorentz symmetry drastically simplifies the computation when we study the perturbation theory with respect to $\mu$. Since the $K$ charge is conserved and the perturbed operator is only positively charged under $K$, a single term out of infinite perturbative series give non-zero results in the perturbative computation of the correlation functions.

\section{Holographic very special $T\bar{J}$ deformation}

\subsection{double trace deformation}
In  holographic descriptions of large $N$ conformal field theories, a bulk supergravity field directly corresponds to a single trace operator of the dual conformal field theories. The single trace deformation of the conformal field theory therefore corresponds to a change of the bulk background. In contrast, in the double trace deformation we do not change the bulk background, but we change the boundary conditions of the supergravity fields \cite{Gubser:2002vv,Bzowski:2018pcy}. In the holographic description of the double trace deformations corresponding to the very special $T\bar{J}$ deformations, we therefore change the boundary conditions for the metric and a bulk vector field so that they break the full isometry of the AdS space-time but preserves the twisted Lorentz symmetry.

We will focus on the effective bulk theory of the  Einstein gravity coupled with $SL(2,\mathbb{R})$ Chern-Smions theory. The ingredients are more or less minimal in the sense that it describes the energy-momentum tensor and the $SL(2,\mathbb{R})$ chiral current algebra so that the very special $T\bar{J}$ deformations based on the twisted Lorentz symmetry are possible. 
The bulk action is given by 
\begin{align}
 S = \frac{k}{4\pi} \int d^3x \sqrt{|g|} (R + 2) +  \frac{\tilde{k}}{4\pi} \int \mathrm{Tr}\left(AdA + \frac{2}{3} A^3 \right) \ . \label{bulka}
\end{align}
Here the Chern-Simons gauge field $A$ takes the value in generators of $SL(2,\mathbb{R})$ (i.e. $A = A^{(i)}_M T^{(i)} dx^M$ with $i = +,0,-$) and the trace is over the  $SL(2,\mathbb{R})$ algebra. In the asymptotic AdS space-time with the standard Dirichlet boundary condition, the bulk action \eqref{bulka} describes the holographic dual of a conformal field theory with $SL(2,\mathbb{R})$ current algebra. Although the classical equations of motion do not depend on $k$ and $\tilde{k}$, we will implicitly assume large $k$ and large $\tilde{k}$ limit to suppress the quantum corrections.

In order to make a holographic interpretation of the bulk theory, we have to specify the boundary condition. Let us first consider the dual description of the undeformed conformal field theory.
For the metric we choose the Fefferman-Graham gauge
\begin{align}
ds^2 = \frac{dz^2}{z^2} + \left(\frac{g_{\mu\nu}^{(0)}}{z^2} + g_{\mu\nu}^{(2)} + \cdots \right) dx^\mu dx^\nu
\end{align}
with the fixed boundary metric $g_{\mu\nu}^{(0)}$ which is identified with the source for the energy-momentum tensor in the dual conformal field theory. 

For the bulk gauge field, we would like to have a holographic realization of right-moving $SL(2,\mathbb{R})$ current algebra (in the undeformed theory). For this purpose, we first put the boundary term 
\begin{align}
S_{\mathrm{boundary}} = -\frac{\tilde{k}}{16\pi} \int d^2x \sqrt{|\gamma|}\gamma^{\mu\nu} \mathrm{Tr}(A_{\mu} A_\nu) 
\end{align}
to make the variation principle well-defined. Then we choose the Dirichlet boundary condition $\lim_{z \to 0} A_{-} = a_{-}$. Here $a_{-}$ will be identified with the source for $J^-$ (or $\bar{J}$ in the holomorphic coordinate) of the dual conformal field theory. We will use the radial gauge $A_z = 0$, and the remaining component $A_{+}$, which will be identified with $J^-$ of the dual conformal field theory, is determined from the bulk equations of motion.


In \cite{Bzowski:2018pcy}, they proposed the boundary condition that corresponds to the $T\bar{J}$ deformations. In our case, we take the $U(1)$ current $\bar{J}$ as the null current $J^{-(+)}$ out of the three generators of $SL(2,\mathbb{R})$. Suppose, before the deformation, we have a recipe to compute the partition function $Z[e,a]$ with respect to the undeformed source e.g. two-dimensional vielbein $e_{a}^{\alpha}$ and the two-dimensional  background $ SL(2,\mathbb{R})$ gauge field $a_{\alpha}^{(i)}$.\footnote{We use the convention that $\alpha, \beta \cdots$ refer to the two-dimensional space-time indices, while $a, b \cdots$ are local Lorentz indices. In the bulk, we use $M,N, \cdots$ and $A,B, \cdots$.} 
 Then, in the very $T\bar{J}$ deformed theory, the new boundary condition becomes
\begin{align}
e_a^{\alpha} &= \tilde{e}_a^{\alpha} + \mu_a J^{\alpha (+)} \cr
a^{(+)}_\alpha &= \tilde{a}^{(+)}_{\alpha} + \mu_a T^{a}_{\alpha} \cr 
a_{\alpha}^{(0)} &= \tilde{a}_{\alpha}^{(0)} \ ,  \cr
a^{(-)}_\alpha  &= \tilde{a}^{(-)}_{\alpha} \label{newb}
\end{align}
with $\mu_- = \mu$ and $\mu_+=0$ corresponding to the $T\bar{J}$ deformation.

Here, tilted quantities are sources in the very special $T\bar{J}$ deformed theory. On the other hand, $J^{\alpha(i)}$ and $T^{a}_{\alpha}$ are expectation values of the current and the energy-momentum tensor (in the original theory), so once we somehow know the original partition function $Z[e,a]$, this gives a (possibly non-linear) relation between the  expectation values of $J^{\alpha(i)}$ and $T^{a}_{\alpha}$ and the source  $\tilde{e}_a^{\alpha}$ and $\tilde{a}_{\alpha}^{(i)}$ (or $e_a^\alpha$ and $a_\alpha^{(i)}$), which can be, at least in principle, solved. 

In the holographic background in the large $N$ limit (or large $k$ limit), we compute $Z[e,a]$ by using the standard GKP-Witten prescription with the Dirichlet boundary condition for the metric and the gauge field.
Since $T^{a}_{\alpha}$ and $J^{\alpha(i)}$ are computed from the holography by deriving the on-shell action with respect to  $e_a^{\alpha}$ and $a_{\alpha}^{(i)}$, the boundary condition employed here is essentially the Robin-type boundary condition in the asymptotically AdS space-time. In the large $\mu$ limit, it effectively becomes the Neumann boundary condition.

Eventually, we are going to set $\tilde{e}_a^{\alpha} = \delta^\alpha_a$ and $\tilde{a}_{\alpha}^{(i)} = 0 $ to describe the flat background in the very special $T\bar{J}$ deformed theory. Let us see how the boundary condition breaks the Lorentz symmetry while it preserves the twisted Lorentz symmetry in this particular background. Under the Lorentz transformation $J^{-(+)}$ is charged, so the first two conditions in \eqref{newb} break the Lorentz symmetry. However, if we define the twisted Lorentz symmetry as the sum of the original Lorentz transformation and the Cartan of $SL(2,\mathbb{R})$ i.e. $J^{(0)}$ as $J_{\mathrm{twisted}} = L_0 - \bar{L}_0 + J^{(0)}$, then $J^{-(+)}$ becomes neutral under the twisted deformation. In other words, the new boundary condition \eqref{newb} with $\tilde{e}_{\alpha} = \delta^\alpha_a$ and $\tilde{a}_{\alpha}^{(i)} = 0 $ is invariant under the twisted Lorentz transformation, so that the entire formalism preserves the  twisted Lorentz symmetry. 

One may also study the conservation of the holographic energy-momentum tensor as in \cite{Bzowski:2018pcy}. With  with $\tilde{e}_{\alpha} = \delta^\alpha_a$ and $\tilde{a}_{\alpha} = 0$, the two most important equations are
\begin{align}
\partial_+ T^{+}_{\ -} + \partial_- (\mu J^{-(+)} T^{+}_{\ -}) &= 0 
\end{align}
and 
\begin{align}
\partial_- J^{-(0)} = \mu J^{-(+)} T^{+}_{\ -} \ , \label{conserv}
\end{align}
which enable us to verify the conservation of the twisted Lorentz current \cite{Nakayama:2018ujt}
\begin{align}
\partial_+(x^- T^{+}_{\ -}) + \partial_- (x^- T^{-}_{\ -} - J^{-(0)}) = 0 \ .
\end{align}
The latter equation \eqref{conserv} is a consequence of the Chern-Simons equations of motion 
\begin{align}
\partial_+ A^{(0)}_- - \partial_- A^{(0)}_+ = A_+^{(+)}A_-^{(-)} - A_-^{(+)}A_+
^{(-)}  \ .
\end{align}
Without the source in the very special $T\bar{J}$ deformed conformal field theory, $\mathrm{lim}_{z \to 0} A^{(0)}_- = 0$ and  $\mathrm{lim}_{z \to 0} A^{(-)}_- = 0$, but $\mathrm{lim}_{z \to 0} A^{(+)}_- = \mu T^{-}_{-}$, so we obtain \eqref{conserv} by noting $\mathrm{lim}_{z \to 0} A_+ = J_+ = -J^{-}$.

There is an alternative formulation of the gravity part by using the $SL(2,\mathbb{R})\times SL(2,\mathbb{R})$ Chern-Simons theory instead of the Einstein gravity \cite{Witten:1988hc}. While the formulation is completely equivalent to the Einstein gravity with the cosmological constant at the classical level,  the reformulation may be of theoretical beauty.\footnote{There has been an on-going discussion if the Chern-Simons formulation may make sense at the quantum level with the holographic interpretation, which we will not address in this paper.}
We are going to replace the Einstein-Hilbert action with the two Chern-Simons action
\begin{align}
S_{\mathrm{gravity}} = \frac{k}{4\pi} \int \mathrm{Tr} \left(\mathcal{A} d\mathcal{A} + \frac{2}{3} \mathcal{AAA} \right) - \frac{k}{4\pi} \int \mathrm{Tr} \left(\bar{\mathcal{A}} d\bar{\mathcal{A}} + \frac{2}{3}  \mathcal{\bar{A}\bar{A}\bar{A}} \right) -  \frac{k}{4\pi} \int d\mathrm{Tr} \left( \mathcal{A \bar{A}} \right)
\end{align}
where the Chern-Simons gauge fields are related to the three-dimensional vielbein and spin connection as
\begin{align}
\mathcal{A}_M &= (w_M^{(i)} + e_M^{(i)}) T^{(i)}  \cr
\mathcal{\bar{A}}_M &= (w_M^{(i)} - e_M^{(i)}) T^{(i)} \ ,
\end{align}
where we have identified the $SL(2,\mathbb{R})$ indices with the local Lorentz indices. 
The bulk metric is therefore given by
\begin{align}
g_{MN} = \frac{1}{2}\mathrm{Tr}((\mathcal{A}-\bar{\mathcal{A}})_M (\mathcal{A}-\bar{\mathcal{A}})_N) \ .
\end{align}

In order to construct the holographic dictionary, we need to specify the boundary conditions. 
The Chern-Simons gauge field in the Fefferman-Graham gauge \cite{Coussaert:1995zp,Banados:1998gg} corresponds to
\begin{align}
\mathcal{A}_z &= -z^{-2}{b^{-1} \partial_z b} \cr 
\mathcal{A}_{\mu} &= b^{-1} a_{\mu}(x^\pm) b  \cr
\bar{\mathcal{A}}_z &= -z^{-2} \bar{b}^{-1} \partial_z \bar{b} \cr
 \bar{\mathcal{A}}_{\mu} &= \bar{b}^{-1} \bar{a}_\mu(x^{\pm}) \bar{b} 
\end{align}
with $b = e^{-(\log z) T^{(0)}}$, $\bar{b} = b^{-1}$, where $T^{(0)}$ is the Cartan of the $SL(2,\mathbb{R})$ generator. We also supplement the condition \cite{Banados:2002ey,Banados:2004nr} 
\begin{align}
\mathrm{Tr}[(\mathcal{A-}\bar{\mathcal{A}})_\mu T^{(0)}] = 0 \ .
\end{align}
One may directly check that this ansatz leads to the Graham-Fefferman form of the three-dimensional metric ansatz.

Now in order to discuss the holographic description of the very special $T\bar{J}$ deformations in the Chern-Simons gravity, we first introduce another $SL(2,\mathbb{R})$ Chern-Simons theory in the bulk whose action and the (undeformed) boundary conditions are specified as before. Then we introduce the deformed boundary condition 
\begin{align}
e_a^{\alpha} &= \tilde{e}_a^{\alpha} + \mu_a J^{\alpha (+)} \cr
a^{(+)}_\alpha &= \tilde{a}^{(+)}_{\alpha} + \mu_a T^{a}_{\alpha} \cr 
a_{\alpha}^{(0)} &= \tilde{a}_{\alpha}^{(0)} \ ,  \cr
a^{(-)}_\alpha  &= \tilde{a}^{(-)}_{\alpha} \label{newbbb}
\end{align}
with $\mu_- = \mu$ and $\mu_+=0$ corresponding to the very special $T\bar{J}$ deformation.

We will not analyze how this boundary conditions lead to the symmetry that is compatible with the very special $T\bar{J}$ deformations because it is classically identical to the Einstein formulation.
Note that although the three $SL(2,\mathbb{R})$ Chern-Simons theories are more or less identical in the bulk (except for the choice of the level), the boundary condition makes them behave differently.

\subsection{single trace deformation}
Let us now consider a holographic description of the single trace very special $T\bar{J}$ deformations. As emphasized in the introduction, it is not always guaranteed that a given (holographic) conformal field theory with a $SL(2,\mathbb{R})$ current algebra possesses the single trace operator with the same quantum number as the double trace $T\bar{J}$ operator. For this purpose, we need a further assumption that the bulk theory has a vector field whose scaling dimension is $(h,\bar{h}) = (2,1)$. It is typical that such a vector field is given by a Kaluza-Klein tower of graviton and Kalb-Ramond field as we will see.

Let us take the background $AdS_3$ and an ``internal space" $M$ with $SL(2,\mathbb{R})$ isometry as a starting point before the $T\bar{J}$ deformation. For instance, we will discuss the case with $M = AdS_3$ or $H_3^+$. The bulk $AdS_3$ may admit a worldsheet string realization with the NS-NS flux. Then the AdS space-time is a classical solution of the Einstein equation and may be embedded in the full string theory (once the central charge is properly chosen).  More precisely, the worldsheet sigma model is given by
\begin{align}
S = \int d^2 w 2k \left(\frac{\partial z \bar{\partial} z -2 \partial {x}^+  \bar{\partial} x^-}{z^2} \right) \ ,
\end{align}
where the target space metric and the Kalb-Ramond field is 
\begin{align}
ds^2 &=  2k \frac{dz^2 -2dx^+ dx^-}{z^2} \cr
B &= -4k \frac{dx^+ dx^-}{z^2}  \ .
\end{align}
The worldsheet theory is conformal invariant and can be a part of the full string theory background. Here the level $k$ determines the size of the AdS space-time.

This bulk geometry has the isometry of $SO(2,3)$ generated by the translation in $x^+$ and $x^-$, the Lorentz transformation rotating $x^{\pm}$, dilatation $(z,x^{\pm}) \to \lambda (z, x^{\pm})$ as well as the special conformal transformation 
\begin{align}
\delta x^+ &= -z^2 \cr
\delta x^- &= -2(x^-)^2  \cr
\delta z  &= -2(x^-)z 
\end{align}
and 
\begin{align}
\delta x^+ &= -2(x^+)^2 \cr
\delta x^- &= -z^2 \cr 
\delta z  &= -2(x^+)z \ .
\end{align}

Correspondingly, the worldsheet theory has the $SL(2,\mathbb{R}) \times SL(2,\mathbb{R})$ current algebra \cite{Chakraborty:2018vja,Apolo:2018qpq}  whose left-moving part is 
\begin{align}
j^+ &= -k\frac{\sqrt{2}}{z^2}\partial x^- \cr
j^0 &= -k\left(-\frac{2}{z^2} x^+ \partial x^- + \frac{1}{z} \partial z \right) \cr
j^{-}  &= -k\left(\frac{2\sqrt{2}}{z^2} (x^+)^2 \partial x^- - \frac{2\sqrt{2}}{z} x^+ \partial z + \sqrt{2} \partial x^+ \right)
\end{align}
and similarly for the right-mover. The operator product expansion is given by
\begin{align}
j^+(w) j^- (0) &= \frac{k}{w^2} + \frac{2j^3(0)}{w} \cr
j^3(w) j^3 (0) & = - \frac{k/2}{w^2} \cr
j^3(w) j^{\pm} (0) & = \pm \frac{j^{\pm}(0)}{w} \ . 
\end{align}
We have used lower script to refer to the worldsheet current algebra (rather than the current algebra of the dual conformal field theory) and used $w$ (rather than $z$) for the worldsheet coordinate to avoid the possible confusion.

In order to realize the very special single trace $T\bar{J}$ deformations, we need an additional assumption for the worldsheet theory describing the internal space $M$. Let us suppose that the internal space world-sheet conformal field theory has another right-moving $SL(2,\mathbb{R})$ current algebra $\bar{k}^{i}$ with the OPE
\begin{align}
\bar{k}^+(\bar{w}) \bar{k}^- (0) &= \frac{k}{\bar{w}^2} + \frac{2\bar{k}^3(0)}{\bar{w}} \cr
\bar{k}^3(\bar{w}) \bar{k}^3 (0) & = - \frac{k/2}{\bar{w}^2} \cr
\bar{k}^3(\bar{w}) \bar{k}^{\pm} (0) & = \pm \frac{\bar{k}^{\pm}(0)}{\bar{w}} \ . 
\end{align}
Following the idea of the single trace $T\bar{J}$ deformations realized in string theory \cite{Chakraborty:2018vja,Apolo:2018qpq}, we propose that the worldsheet description of the single trace very special $T\bar{J}$ deformation is  obtained by adding the worldsheet marginal current-current deformations of
\begin{align}
\delta S = \mu  \int d^2w  j^+ \bar{k} \ .
\end{align}
The deformations break the worldsheet symmetry of $SL(2,\mathbb{R}) \times SL(2,\mathbb{R}) \times SL(2,\mathbb{R})$ down to $SL(2,\mathbb{R}) \times  \mathrm{diag}(SL(2,\mathbb{R}) \times SL(2,\mathbb{R})) $ (in addition to the Virasoro symmetries that are always preserved). In particular, although the original target space Lorentz symmetry (rotations of $x^{\pm}$) is broken, the twisted target space Lorentz symmetry is preserved. 

Let us study explicit realizations of $M$. Let us first take the case with $M = AdS_3$. Then the wordsheet sigma model is 
\begin{align}
S = \int d^2w 2\tilde{k} \left(\frac{\partial \tilde{z} \bar{\partial} \tilde{z} -2 \partial {\tilde{x}}^+  \bar{\partial} \tilde{x}^-}{\tilde{z}^2} \right) \ ,
\end{align}
with
\begin{align}
\bar{k}^+ &= -\tilde{k}\frac{1}{\tilde{z}^2}\bar{\partial} \tilde{x}^+ \cr
\bar{k}^0 &= -\tilde{k}\left(\frac{1}{\tilde{z}^2} \tilde{x}^- \bar{\partial} \tilde{x}^+ + \frac{1}{\tilde{z}} \bar{\partial} \tilde{z} \right) \cr
\bar{k}^{-}  &= -\tilde{k}\left(\frac{1}{\tilde{z}^2} (\tilde{x}^-)^2 \bar{\partial} \tilde{x}^+ + \frac{2}{\tilde{z}} \tilde{x}^- \bar{\partial} \tilde{z} - \bar{\partial} \tilde{x}^- \right) \ ,
\end{align}
which generates the worldsheet $SL(2,\mathbb{R})$ current algebra. 
Now, the worldsheet current-current deformation is given by 
\begin{align}
\delta S = \mu \int d^2x \left(\frac{\partial x^+ \bar{\partial} \tilde{x}^+}{z^2 \tilde{z}^2} \right) \ .
\end{align}
The space-time interpretation is that the metric is deformed 
\begin{align}
ds^2 = 2k \frac{dz^2 + dx^+ dx^{-}}{z^2} + 2\tilde{k}\frac{d\tilde{z}^2 + d\tilde{x}^+ d\tilde{x}^-}{\tilde{z}^2} + \mu \frac{dx^+ d\tilde{x}^+}{z^2 \tilde{z}^2} 
\end{align}
and the Kalb-Ramond field is deformed
\begin{align}
B = 2k \frac{dx^+ dx^{-}}{z^2} + 2\tilde{k}\frac{d\tilde{x}^+d\tilde{x}^-}{\tilde{z}^2} + \mu \frac{dx^+ d\tilde{x}^+}{z^2 \tilde{z}^2} \ . 
\end{align}
It is immediate to see that the background breaks the original Lorentz symmetry but preserves the twisted one given by $(x^+,x^-,\tilde{x}^+,\tilde{x}^-) \to ( \lambda x^+, \lambda^{-1} x^-,\lambda^{-1} \tilde{x}^+, \lambda^{} \tilde{x}^-)$.

The case with the $H_3^+$ is more intricate. Let us begin with the $H_3^+$ sigma model
\begin{align}
S = \int d^2x 2\tilde{k} \left(\frac{\partial \tilde{z} \bar{\partial} \tilde{z} + \partial {\xi} \bar{\partial} \bar{\xi}^-}{\tilde{z}^2} \right) \ ,
\end{align}
where $\xi = \tilde{x}_1+i\tilde{x}_2$ and $\bar{\xi} = \tilde{x}_1-i\tilde{x}_2$ are complex conjugate. Therefore the model has the pure imaginary NS-NS flux
\begin{align}
B = i \frac{d\tilde{x}_1 d\tilde{x}_2}{z^2} \ .
\end{align}
One may perform the same current-current deformations as in the $AdS_3$ case with the complex metric (or more precisely non-real metric), but the more intricate choice of the deformation
\begin{align}
\delta S = \mu \int d^2x \left(\frac{\partial x^+ \bar{\partial}w + \partial \bar{w} \bar{\partial} x^+}{z^2 \bar{z}^2} \right) \ 
\end{align}
might look better because the metric and the Kalb-Ramond field satisfies the same ``reality condition" as the undeformed theory. This is the combination of two different worldsheet current-current deformations of $j\bar{k}$ and $k\bar{j}$, and it is not immediately obvious if this gives the exact marginal deformations. Since the AdS current part is taken to be the same and null, the first order perturbation is marginal, so it at least passes the first order check.

\section{Discussions}
In this paper, we have proposed holographic descriptions of very special $T\bar{J}$ deformations of conformal field theories. The very special $T\bar{J}$ deformation is peculiar: it preserves ``twisted" Poincar\'e symmetry as well as dilatation, but it has only chiral conformal transformation. Typically, unitary scale invariant relativistic field theories have the full conformal symmetry \cite{Polchinski:1987dy,Hofman:2011zj}, but this is avoided because the existence of a non-compact symmetry and use of the topological twist resulted in non-unitary quantum field theories.

Our holographic construction is classical and it is an interesting question if the gravity side can be quantized e.g. in the full string setup. One of our formulation for the double trace $T\bar{J}$ deformations is based on the $SL(2,\mathbb{R})\times SL(2,\mathbb{R}) \times SL(2,\mathbb{R})$ Chern-Simons theory, and without our complicated boundary conditions, the quantization may be straight-forward. However, whether the quantum gravity makes sense in particular with our mixed boundary conditions is a different story and it deserves a further study. With this regard, in our discussions, we have restricted ourselves to the parity preserving case with the same Chern-Simons levels for the gravity part, but we may choose three levels of the Chern-Simons theory completely differently. Such choices may be fine with the trivial background source but may cause anomalies in the general background. 

As for the single trace deformation, the resultant deformed background is similar to the one studied in the $T\bar{J}$ deformations, but in our case, the internal space is also non-compact so the further Kaluza-Klein reduction to obtain the warped AdS space-time is not possible. Also one of our explicit examples has two ``time-like" directions in the target space and the physical significance should be carefully studied beyond the formal analysis done in this paper.

\section*{Acknowledgements}
This work is in part supported by JSPS KAKENHI Grant Number 17K14301.


\end{document}